# First in-situ detection of the CN radical in comets and evidence for a distributed source

Nora Hänni[1,*], Kathrin Altwegg[1], Boris Pestoni[1], Martin Rubin[1], Isaac Schroeder[1], Markus Schuhmann[1], Susanne Wampfler[2]

[1] Physikalisches Institut, University of Bern, Sidlerstrasse 5, CH-3012 Bern, Switzerland
[2] Center for Space and Habitability, University of Bern, Gesellschaftsstrasse 6, CH-3012 Bern, Switzerland
* Correspondence to E-mail: nora.haenni@space.unibe.ch; Tel.: +41 31 631 4449.

## Abstract
Although the debate regarding the origin of the cyano (CN) radical in comets has been ongoing for many decades, it has yielded no definitive answer to date. CN could previously only be studied remotely, strongly hampering efforts to constrain its origin because of very limited spatial information. Thanks to the European Space Agency's Rosetta spacecraft, which orbited comet 67P/Churyumov-Gerasimenko for two years, we can investigate, for the first time, CN around a comet at high spatial and temporal resolution. On board Rosetta's orbiter module, the high-resolution double-focusing mass spectrometer DFMS, part of the ROSINA instrument suite, analyzed the neutral volatiles (including HCN and the CN radical) in the inner coma of the comet throughout that whole two-year phase and at variable cometocentric distances. From a thorough analysis of the full-mission data, the abundance of CN radicals in the cometary coma has been derived. Data from a close flyby event in February 2015 indicate a distributed origin for the CN radical in comet 67P/Churyumov-Gerasimenko.

## 1. INTRODUCTION
Until 1984 it was widely believed in the cometary science community that hydrogen cyanide (HCN) was the sole parent of the cometary CN radical. CN is a photodissociation product of HCN with a quantum yield of 0.97, cf. Huebner et al. (1992). Bockelée-Morvan et al. (1984) first questioned this notion based on the remote observations of comet C/1983 H1 (IRAS-Araki-Alcock). The authors deduced an upper limit for the HCN production rate which was smaller than the CN production rate previously derived by A'Hearn et al. (1983). This puzzling observation raised many questions regarding the origin of the CN radical, which remains a riddle even today. Bockelée-Morvan et al. (1985) also demonstrated that the destruction scale length of the CN parent according to the Haser model (Haser (1957)) could not be associated with HCN photodissociation alone. Not only would production rates and scale lengths of parent species and CN have to be compatible, the strongly collimated CN jets observed for the first time in comet 1P/Halley by A'Hearn et al. (1986) would also have to be explained. Even though their results showed a jet morphology consistent with a parent-daughter relation between HCN and CN, Woodney et al. (2002) pointed out discrepancies between the destruction scale length of HCN and the formation scale length of CN. A typical HCN destruction scale length is about 80,000 km (based on photodestruction rates by Huebner et al. (1992) for a heliocentric distance of 1 au, quiet Sun conditions and an assumed gas velocity of 1 km/s). Fray et al. (2005) reviewed many remote observations of HCN and CN in comets and showed that the CN formation scale length is only on the order of 25,000 km at 1 au. In particular, at heliocentric distances smaller than 3 au, all eight of the comets reviewed exhibited CN formation scale lengths that were lower than the HCN destruction scale length. Regarding the CN and HCN production rates, it seems that there are comets (1P/Halley, C/1990 K1 (Levy), C/1996 B2 (Hyakutake) and C/1995 O1 (Hale-Bopp)) in which HCN production is similar enough to the CN production to sufficiently explain it, but others (C/1989 X1 (Austin), C/1983 H1 (IRAS-Araki-Alcock), 21P/Giacobini-Zinner and 107P/Wilson-Harrington) where that is not the case. Besides CN-bearing volatile species, the literature also discusses solids degraded thermally or by UV/cosmic rays as additional potential parents. The following paragraphs provide a brief overview on the two main candidate groups of parent species:

1.) CN-bearing refractories ejected with dust particles from the cometary nucleus could yield HCN or even CN directly while travelling along their outbound trajectories. These species are therefore referred to as distributed or extended sources. In the last few decades, research has focused mainly on four different compounds:

- HCN polymers have been discussed as candidates for the black refractory matter on comets e.g. by Matthews (2006). Fray (2004) and Fray et al. (2004) studied their degradation under the influence of temperature and UV irradiation. Their HCN production was small, as these polymers carbonize when undergoing pyrolysis, with the major degradation product being ammonia ($NH_3$). Irradiation produced HCN even less efficiently than thermal degradation. At present, detections of HCN polymers in space are also only tentative.
- Hexamethylenetetramine (HMT) has not yet been identified in comets or the interstellar medium, but it can be synthesized in interstellar ice analogs, cf. Bernstein et al. (1995). Fray et al. (2004) reported production of HCN at elevated temperatures or UV irradiation. For both HCN polymers and HMT, it is not clear to the authors whether CN radicals are also a product of the degradation processes and additional laboratory experiments are necessary.
- Tholin (a reddish mixture of carbon-, nitrogen- and hydrogen-containing complex organic molecules named and proposed as constituent of interstellar grains and gas by Sagan et al. (1979)) may also produce CN radicals upon degradation, the simplest route being via nitrile groups. Such groups were suggested as abundant constituents e.g. by Imanaka et al. (2004) based on Fourier-transformed infrared data and reported to mainly occur in tholins that were synthesized at high pressures. Their presence was confirmed in $^{13}C$ and $^{15}N$ solid state nuclear magnetic resonance experiments by Derenne et al. (2012). Already Ehrenfreund et al. (1995) and Pietrogrande et al. (2001) observed various species from tholin pyrolysis experiments, HCN and small nitriles apparently being very common. However, laboratory experiments to investigate the possibility of direct CN production from tholins seem to be outstanding to date.
- CHON grains are cometary dust particles composed of the elements carbon (C), hydrogen (H), oxygen (O) and nitrogen (N). Lawler and Brownlee (1992) showed in an analysis of Vega-1 and Giotto data collected during their respective flybys of comet Halley that the refractory CHON component was interspersed with silicate components at sub-micrometer scales. They proposed that, due to sublimation, many particles have a common proportion of CHON and silicate material and disagreed with earlier analyses by Jessberger et al. (1988) wherein various compositional groups were identified. It can be assumed that CN could (a.) directly photodissociate from molecules on the surface of the grain, (b.) photosputter off the surface or (c.) have a parent which can sublimate from the grain and subsequently photodissociate after a short lifetime of less than about 1000 s, cf. A'Hearn et al. (1986) and Combi (1987).

2.) Volatile CN-bearing species have been investigated in various remote observation studies reviewed by Fray et al. (2005). Like HCN, such species produce CN radicals upon photodissociation. The yield of this process depends mainly on the photodestruction rate of the candidate molecule and on the corresponding quantum yield. According to Fray et al. (2005), cyanogen ($C_2N_2$), cyanoacetylene ($HC_3N$) and acetonitrile ($CH_3CN$) are the most promising candidates aside from HCN itself. Cyanoacetylene and acetonitrile are known to be present in trace amounts in comets as reported for comet Hale-Bopp by Bockelée-Morvan et al. (2000) and for comet 67P/Churyumov-Gerasimenko by Le Roy et al. (2015) and Rubin et al. (2019), while cyanogen has not been detected in comets prior to the European Space Agency's (ESA) Rosetta mission. Altwegg et al. (2019) were the first to report single instances where cyanogen was observed in the inner coma comet 67P/Churyumov-Gerasimenko.

This work is dedicated to the first in-situ study of the cometary CN radical, based on data collected by the Rosetta Orbiter Spectrometer for Ion and Neutral Analysis (ROSINA; Balsiger et al. (2007)) from the inner coma of a comet, namely that of 67P/Churyumov-Gerasimenko (hereafter 67P). Rosetta was launched by ESA to accompany 67P for two years as it passed through its perihelion, providing data with unprecedented temporal and spatial resolution. The experimental methodology used to analyze the high-resolution mass spectra is described in detail, following which the data are presented and correlations with acquisition parameters, especially the cometocentric distance, are discussed. Finally, the feasibility of various possible origins of CN is considered and suggestions for follow-up investigations are proposed.

## 2. EXPERIMENTAL METHODOLOGY
### 2.1. Instrumentation
The Double Focusing Mass Spectrometer (DFMS) was developed as part of the ROSINA instrument package for the orbiter module of ESA's Rosetta mission. The DFMS is a sector field electron-impact ionization mass spectrometer (EI-MS) constructed in the Mattauch-Herzog configuration, cf. Mattauch et al. (1934). It uses 45 eV electrons to ionize neutral volatiles in the inner coma of 67P. In the following analysis, the positive charge produced under EI has been omitted to simplify the notation. The ionization process often causes fragmentation of the specimen. The ionized molecules, as well as the charged fragments, are subsequently transferred through an electrostatic analyzer and separated according to their mass-to-charge ratios $m/z$ in the electric and magnetic fields before finally impacting on the stack of two Micro Channel Plates (MCPs) mounted in a Chevron configuration. The MCPs release a cascade of electrons upon impact and the charge is collected on two rows (A and B) of Linear Electron Detector Array (LEDA) anodes with 512 pixels each. By adjusting the voltage across the MCP for each $m/z$ value individually, a suitable amplification (i.e. gain) is attained and saturation avoided. In this way, the gain applied in each measurement could be varied and a high dynamic range of the order of $10^{10}$ was achieved. The DFMS has a high mass resolution of m/Δm ≈ 3000 (at 1 per cent peak height for *m/z* = 28 u/e) and covers an *m/z* range from 12 to 180 u/e. More detailed information may be found in Balsiger et al. (2007).

## 2.2. Data collection

The DFMS was designed to detect both cometary neutrals and ions. Data analyzed for this work were collected by operating the DFMS in one of the pre-programmed high-resolution neutral gas modes, which scan a pre-defined *m/z* range. Each integer *m/z* value is scanned individually, with an integration time of 20 s per spectrum. Two additional measurements of *m/z* = 18 u/e (water) are always included at the beginning and end of each mode. Together with the 10 s between each m/z value required for setting the appropriate voltages, a typical scan across the mass range from *m/z* = 13 to 100 u/e takes about 45 min. To repel ambient cations when operating in the neutral gas modes, a positive voltage is applied to the ion suppressor plate around the entrance to the ionization source box. Ambient anions are accelerated toward the plate but ultimately do not possess suitable energy and charge for transmission through the instrument. The DFMS generally accepts incident particles from a 20° x 20° field of view and is designed and optimized mainly for the transmission of particles with energies on the order of a few tenths of an eV.

## 2.3. Data evaluation

The data evaluation procedure has been previously described by Rubin et al. (2019) and in the literature cited therein. This work focuses mainly on the CN signal on mass 26 u/e and HCN one on mass 27 u/e. Two representative spectra collected at the beginning of the mission phase on MCP row A are shown in Fig. 1, which also includes a typical double-Gaussian peak fit. The CN and HCN signals are clearly detectable and only slightly overlap with other signals. Minor isotopologues (<1 per cent relative abundance) have been neglected. Fitting errors as well as statistical errors of the signals are small, <5 per cent for HCN. A 15 ~ 20 per cent error on the instrument sensitivity (proportional to $(m/z)^{-0.8}$; Calmonte (2015)) and a 20 per cent error on the detector gain and pixel gain correction have to be considered in addition. Details related to gain and pixel gain corrections may be found in Schroeder et al. (2019). However, these errors at least partially cancel out if signal ratios are considered. A correction for the off-nadir pointing angle θ has also been performed by division with cos(θ), which represents the geometrical cross-section of the DFMS' ion source exposed to the comet. The analysis presented in this work is based on two types of values: $c_X$, where X is a specific chemical species, corresponds to the automatically extracted and corrected detector signals (number of ions per accumulation time) and $n_X$, which corresponds to the local number density derived from $c_X$ by taking into account fragmentation under electron impact inside the mass spectrometer, as well as the sensitivity, which includes ionization cross-section, transmission, and detector yield. To derive the density related to the portion of CN signal which is not explained by fragmentation, we employ the ionization cross-section of the CN radical reported by Pandya et al. (2012). As the electron-impact-induced fragmentation branching of the CN radical according to CN + e$^-$ → C$^+$ + N + 2e$^-$ and CN + e$^-$ → N$^+$ + C + 2e$^-$ is not known, we assume no fragmentation at all, making the derived density a lower limit. Based on the individual errors indicated above, the total error on $c_X$ and $n_X$ is estimated to be 30 per cent. Their ratios, however, have an error <10 per cent.

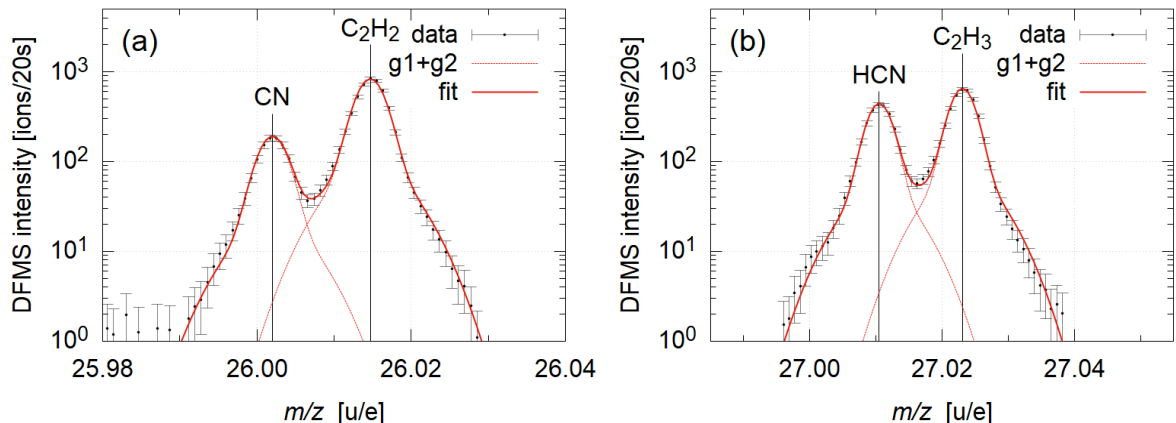

Figure 1. Typical spectra at integer masses $m/z$ = 26 u/e (a) and 27 u/e (b) collected at the beginning of the Rosetta mission phase on 17$^{th}$ September 2014 over the southern hemisphere of comet 67P. Both the CN as well as the HCN signals are clearly separated from the signals of $C_2H_2$ and $C_2H_3$ respectively. Error bars indicate only the statistical error.

## 3. RESULTS
### 3.1. Total CN signal

Any CN observed by the high-resolution mass spectrometer DFMS onboard the Rosetta orbiter module either enters the instrument as a neutral coma species (the CN radical itself is neutral), where it is ionized and subsequently detected, or is formed directly in the ion source from a fragmenting CN-bearing coma species. For the example of HCN, ionization occurs according to HCN + $e^-$ → $HCN^+$ + $2e^-$, while fragmentation occurs according to HCN + $e^-$ → $CN^+$ + H + $2e^-$. Other fragments, namely $C^+$, $CH^+$, $N^+$ and $NH^+$, with intensities of less than 5 per cent relative to the most intense peak of $HCN^+$ according to the NIST (National Institute of Standards and Technology) Standard Reference Database Number 69, are neglected in our analysis. It is important to note that these two competing processes, ionization and fragmentation, yield a constant ratio CN/HCN specific to the ionization energy and instrument /mass-dependent sensitivity. CN radicals, which are ionized into CN ions in the DFMS, contribute to the total CN signal $c_{CNt}$, together with the CN fragments produced by EI from CN-bearing neutral molecules in the cometary coma. Hence, the portion of the CN signal due to the fragmentation inside the ion source $c_{CNf}$ must be subtracted from the total CN signal $c_{CNt}$ to obtain the portion of cometary CN radicals according to $c_{CNr}$ = $c_{CNt}$ - $c_{CNf}$. $c_{CNf}$ contains multiple contributions, e.g. $c_{CNf-HCN}$ from fragmentation of HCN inside the ion source discussed above, $c_{CNf-HC3N}$ from $HC_3N$, and $c_{CNf-HNCO}$ from HNCO among others. (This does not exclude the processes of fragmentation occurring in the coma, which will be discussed later.) Due to their toxicity, HCN and other CN-bearing species observed in the coma of 67P have not been calibrated for in the laboratory. Such tests are normally conducted with a twin model of the DFMS attached to the CAlibration SYstem for the Mass spectrometer Instrument ROSINA (CASYMIR; Westermann et al. (2001)). The fragmentation patterns of such species are therefore not known for the DFMS. Available data from the NIST Standard Reference Database Number 69 and other literature report that the CN signal produced by the fragmentation of HCN ranges in relative intensity values from 0.168 (NIST) to 0.148 (Kusch et al. (1937)) and 0.11 (Stevenson (1950)). When comparing ROSINA/DFMS data to these values, it is crucial to realize that differences must be expected due to the different ionization energies used. The DFMS operates with a 45 eV electron beam, whereas the NIST standard is 70 eV. Lower energy electrons provide less energy for fragmentation and consequently a lower fragment yield is observed. With regard to the $c_{CNt}/c_{HCN}$ ratio we extracted from our data, as previously described in subsection 2.3., this means that lower values can be expected, as compared to NIST.

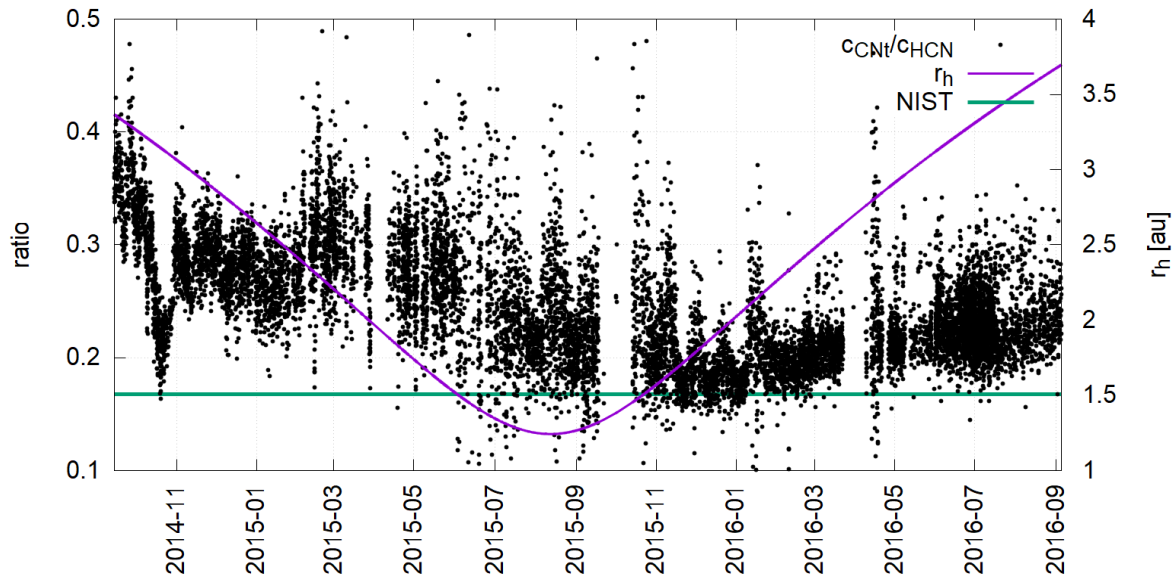

Figure 2. $c_{CNt}/c_{HCN}$ ratio from nadir-corrected MCP row A full-mission data. The green line indicates the expected constant CN signal relative to HCN (NIST value) due to the fragmentation of the latter under EI inside the ion source, i.e. $c_{CNf-HCN}/c_{HCN}$. The purple line indicates 67P's heliocentric distance $r_h$. Error bars on $c_{CNt}/c_{HCN}$ have been omitted for visual clarity.

Fig. 2 shows the $c_{CNt}/c_{HCN}$ ratios observed by ROSINA/DFMS over the course of the whole Rosetta mission phase. Surprisingly, the ratio ranges from values as high as ~ 0.40 at the beginning of the mission (the $c_{CNt}/c_{HCN}$ ratio derived from the spectra shown in Fig. 1 is 0.41) to values as low as ~ 0.15 shortly after perihelion. Such variation cannot be explained solely by the fragmentation of HCN inside the instrument, which is a constant process.

Fig. 3 shows the correlation of $c_{CNt}/c_{HCN}$ with the latitude of the orbiting spacecraft above the cometary surface for three characteristic periods. The top panel is near the beginning of the active mission phase in August 2014, the middle panel around the time of 67P's perihelion in August 2015 and the bottom panel at the end of the mission in September 2016. The observed latitudinal variation of the $c_{CNt}/c_{HCN}$ ratios also makes clear that the fragmentation of HCN inside the instrument cannot be the only contributor to the CN signal. Other contributions with latitudinal distributions different from that of HCN have to be overlaid to reproduce such behavior. The $c_{CNt}/c_{HCN}$ ratio at the beginning of the mission phase, cf. Fig. 3 (top), is high when Rosetta was measuring above the southern hemisphere. During the 10 km-orbital phase in late October 2014, at about 3.15 au from the Sun, the initially clear latitudinal correlation is gradually lost. Around the time of 67P's perihelion, cf. Fig. 3 (middle), the correlation between the $c_{CNt}/c_{HCN}$ ratio and the latitude is the inverse of what was previously observed at the beginning of the mission, with ratios now generally being higher when the spacecraft was passing over the northern latitudes. This inversion reflects seasonal variations, which in turn indicates that solar illumination plays a crucial role. After the inbound equinox in May 2015 (at about 2.2 au from the Sun), the Sun was over the southern hemisphere, causing a short and hot summer there. The outbound equinox in March 2016 (at about 2.7 au from the Sun) changed the illumination conditions yet again, heralding a long but less intense summer over the northern hemisphere. The $c_{CNt}/c_{HCN}$ ratio reaches minimal values of ~ 0.15 (comparable with the aforementioned NIST value of 0.168 and listed in Table 1 below) in the fourth quarter 2015 and hence after the comet's peak activity. This means, $c_{CNt}$ measured after perihelion is clearly governed by fragmentation of HCN. It does not mean, however, that the additional CN source has disappeared. High ratios were frequently observed over the northern hemisphere, where the ratios were also consistently higher in the beginning of the mission (top). This, together with a consistent noise level of about ±0.025 on $c_{CNt}/c_{HCN}$ estimated from the data shown in the top and bottom panels of Fig. 3, indicates a non-instrumental effect which is probably related to dust activity. Towards the end of the mission in September 2016, $c_{CNt}/c_{HCN}$ is higher over the southern latitudes and it increases periodically while the spacecraft is flying southwards, performing its end-of-mission ellipses, cf. Fig. 3 (bottom). Expressed as a function of the heliocentric distance, the slope of the HCN production is steeper than the slope of the production of the additional CN parent species, leading to the small values of $c_{CNt}/c_{HCN}$ around and especially shortly after 67P's perihelion passage. In addition to the variation of $c_{CNt}/c_{HCN}$ over the course of the mission, short-term variations are also visible in the data. However, the underlying forces and mechanisms responsible for these observed trends are strongly interlinked with the origin and production process of the additional CN that cannot be explained by the fragmentation of HCN inside the DFMS alone. In the following paragraph we investigate on the possibility that other cometary CN-bearing volatiles with variable coma abundances contribute CN fragments ($c_{CNf-HCN}$, $c_{CNf-HNCO}$, etc.) to the total amount $c_{CNt}$.

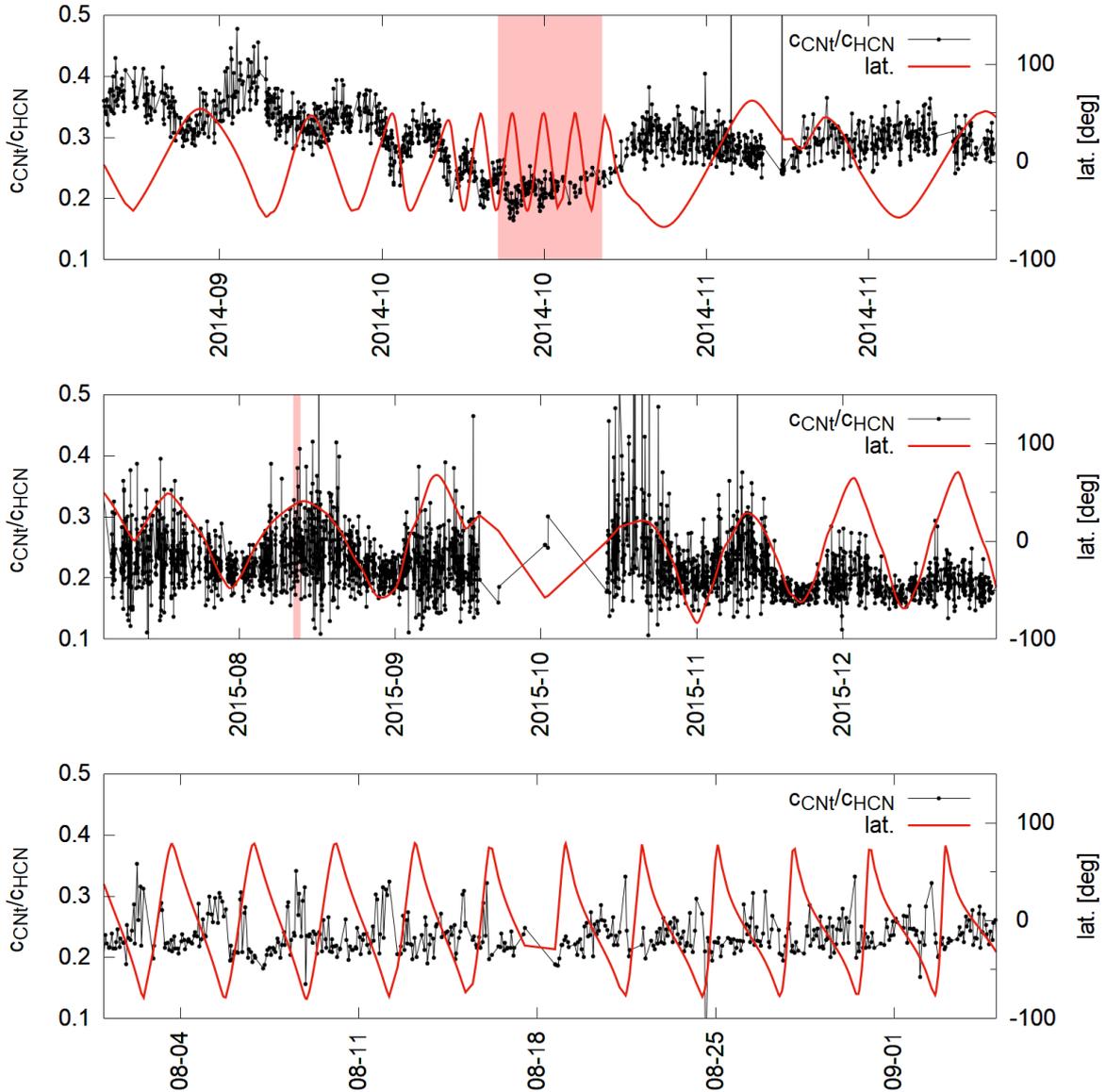

Figure 3. The $c_{CNt}/c_{HCN}$ ratio at the beginning of the Rosetta mission phase (top), around perihelion (middle) and during the end-of-mission elliptical orbits from early August to early September 2016 (bottom). The latitudinal position of the spacecraft above the cometary surface is shown with a red line. The 10 km-orbital phase (15th to 28th October 2014) and the date of the perihelion (13th August 2015) are shaded in pink. 10 per cent error bars on $c_{CNt}/c_{HCN}$ have been omitted for visual clarity.

### 3.2. CN radical contribution

Besides HCN, other volatile CN-bearing species were also observed by ROSINA/DFMS in the coma of comet 67P, cf. Rubin et al. (2019) and Le Roy et al. (2015). All of these species contribute to some extent to the production of CN fragments. Table 1 presents a summary of their abundances (relative to water), together with literature values for CN/parent ratios measured with EI-MS.

Table 1. Abundances (relative to water in percent) of CN-bearing volatiles observed at 67P with ROSINA/DFMS. Bulk abundances were derived from an observational period in May 2015, as explained in Rubin et al. (2019), whereas abundances over the northern (N) and southern (S) hemispheres were derived from single observations in October 2014, as explained in Le Roy et al. (2015). The fragmentation behavior, represented by the $c_{CNf-X}/c_X$ ratio, where X indicates the parent species of the CN fragment, is taken from the referenced literature. All possible isomers of each molecule are listed (since separating them is very challenging in mass spectrometry), beginning with the ones which are the subject of the present fragmentation analysis.

| parent species | chemical structure formula | abundance rel. to water in % | $c_{CNf-X}/c_X$ | Refs. for fragmentation patterns |
|---|---|---|---|---|
| hydrogen cyanide | HCN | bulk: 0.14±0.04 N: 0.09 S: 0.62 | 0.168[+] | NIST |
| hydrogen isocyanide | HNC | | | |
| acetonitrile/methyl cyanide | $CH_3CN$ | bulk: 0.0059±0.0034 N: 0.006 S: 0.016 | 0.016° | NIST/AIST |
| methyl isocyanide | $CH_3NC$ | | 0.020° | NIST: from R. G. Gillis, Aust. Sci. Service |
| isocyanic acid | HNCO | bulk: 0.027±0.016 N: 0.016 S: 0.031 | 0.019 (0.024) | Fischer et al. (2002) (Bogan et al. (1971)) |
| cyanic acid | HOCN | | | |
| formamide | $NH_2COH$ | bulk: 0.0040±0.0023 N: <1e-4 S: <1e-3 | 0.013 | NIST/AIST |
| nitrosomethane | $CH_3NO$ | | 0.029 | NIST |
| formaldehyde oxime | $CH_2NOH$ | | | |
| cyanoacetylene/ propiolonitrile | HCCCN | bulk: 0.00040±0.00023 N: <2e-5 S: <2e-4 | 0.034 | NIST |
| isocyanoacetylene | HCCNC | | | |
| cyanogen | NCCN | bulk: [§] N: - S: - | 0.047 | NIST: from A. A. Kutin, Moscow, Russia |
| Isocyanogen | CNCN | | | |

\* The parent peak is usually also the highest peak in the fragmentation pattern.
[+] Literature specifying other (lower) values for CN/HCN is cited in the main text.
° In unit-resolution NIST data, the signal on mass 26 u/e can be assigned either to CN or to $C_2H_2$ or to a mixture of both. This ambiguity is relevant only for acetonitrile, a molecule with 2 C and 3 H atoms. We assign the reported intensity to the more likely fragment $C_2H_2$. However, attributing it to CN would also not change the overall picture due to the low abundance of acetonitrile.
[§] Rare detections by ROSINA/DFMS during the mission phase (Altwegg, priv. commun.).

Rubin et al. (2019) estimated the volatile bulk inventory of comet 67P based on ROSINA/DFMS data collected over the southern hemisphere in May 2015 (local summer). This time period was selected as being most representative for the cometary bulk material because the comet, then about 1.5 au from the Sun, was already quite active. Closer to perihelion, frequent short-lived outbursts ejected significant quantities of dust, leading to highly variable coma conditions. Rubin et al. (2019) calculated abundances relative to water (in percent). Their data are in fair agreement with the findings of Läuter et al. (submitted to MNRAS), who derived relative production rates as a function of time, as well as the integrated total mass-loss per species over the whole Rosetta mission phase, based on data from ROSINA/DFMS. From the available bulk abundances, which are especially relevant for the time around perihelion, HCN appears to be comet 67P's major CN-bearing species with its bulk abundance of 0.14 per cent relative to water. The CN fragment yield of HCN is expressed as $c_{CNf-HCN}/c_{HCN}$ = 0.168 in Table 1. For HNCO, the species with the second largest bulk abundance, both the bulk abundance and the CN fragment yield are lower than that of HCN by almost an order of magnitude. This rough estimation seems to indicate that only a non-exhaustive portion of the observed CN signal can be attributed to fragmentation of CN-bearing species under EI. Le Roy et al. (2015) derived abundances relative to water from single observations made during the 10 km-orbital phase in late October 2014, at about 3 au from the Sun and over the northern and the southern hemispheres of the comet, respectively. According to their data, all of the CN-bearing species were more abundant, relative to water, over the southern hemisphere (local winter). Water outgassing from the northern hemisphere (local summer) was observed to be 16-times higher than from the southern hemisphere (local winter). Corrected for this factor, HCN outgassing from the north is roughly double the HCN outgassing from the south. As shown in Fig. 2, the higher $c_{CNt}/c_{HCN}$ ratios were observed over the southern (winter) hemisphere and hence not where most of the HCN is originating from.

The residual CN signal $c_{CNr}$ was derived, as explained in subsection 2.3., from the total CN signal $c_{CNt}$ by subtracting the CN produced under EI-fragmentation $c_{CNf}$ according to $c_{CNr} = c_{CNt} - c_{CNf}$, where $c_{CNf} = c_{CNf-HCN} + c_{CNf-HNCO} + \ldots$ The following CN-bearing molecules were taken into account, using the fragmentation patterns referenced in Table 1: hydrogen cyanide (HCN; NIST), acetonitrile ($CH_3CN$; NIST), isocyanic acid (HNCO; Fischer et al. (2002)), formamide ($NH_2COH$; NIST/AIST) and cyanoacetylene ($HC_3N$; NIST and LeRoy priv. commun.). Having thoroughly corrected for the instrumental effects of fragmentation and since externally-generated CN ions cannot enter the instrument during neutral measurement modes for the reasons indicated in subsection 2.2., the only remaining explanation for the residual CN signal $c_{CNr}$ is the neutral CN radical. This result is expected and well in line with remote observations, such as those reviewed by Fray et al. (2005).

After subtraction of the fragmentation contribution of various CN-bearing molecules from $c_{CNt}$, a lower limit for the local number density attributable to the CN radical $n_{CNr}$ can be derived from $c_{CNr}$. Fig. 4 (top) shows $n_{CNr}/n_{HCN}$, the radical's local density relative to the local density of HCN for the full mission phase. The values span several orders of magnitude, ranging from almost 0.3 down to about 0.0001, indicating comparable densities of CN and HCN when 67P was inbound but still far from perihelion and negligible CN densities around the outbound equinox in March 2016. Similar to $c_{CNr}/c_{HCN}$ in Fig. 3, also the number density ratio shows clear latitudinal variations. Both ratios follow the same trends, which is understandable as HCN is the main CN fragment contributor and instrument-inherent fragmentation is a constant process. A modulation of $n_{CNr}/n_{HCN}$ with respect to $c_{CNr}/c_{HCN}$ comes from differences in the ionization cross-sections. Pandya et al. (2012) calculate at 45 eV roughly 2.6 Å$^2$ for CN and 3.3 Å$^2$ for HCN, implying a higher ion yield for HCN than for CN.

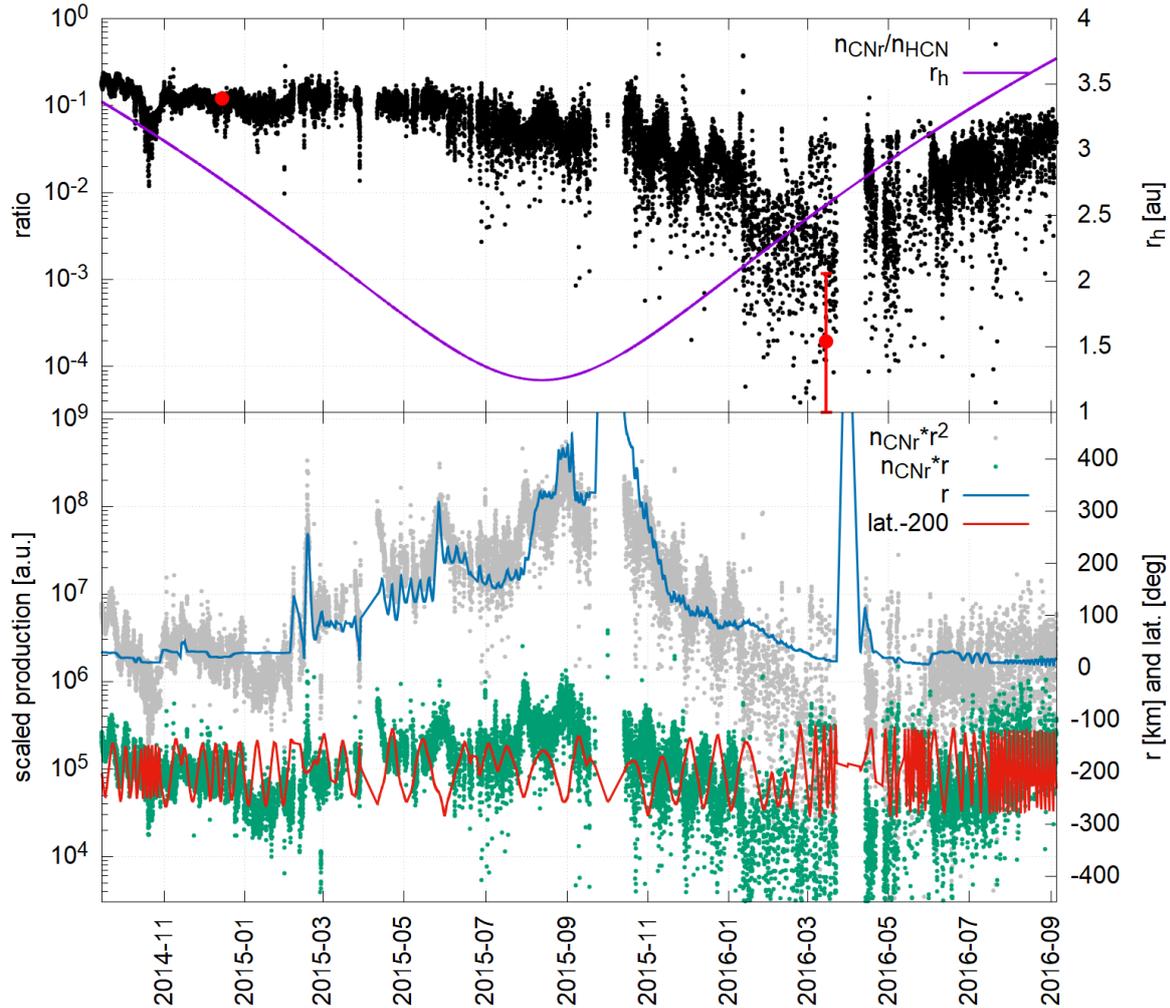

Figure 4. (top) Full mission lower limit local density of CN relative to the local density of HCN as $n_{CNr}/n_{HCN}$ (black dots). The heliocentric distance $r_h$ is indicated as purple line. The relative statistical errors of the density ratio vary between a few per cent (mid-December 2014, where $n_{CNr}$ and $n_{HCN}$ are of the same order of magnitude) and more than 500 per cent (mid-March 2016, where $n_{CNr}$ and $n_{HCN}$ differ by almost four orders of magnitude). Two indicative error bars are shown in red: 4 per cent on 14th December 2014 (smaller than symbol size) and 502 per cent on 15th March 2016. Systematic errors of about 5 per cent must be considered in addition. (bottom) $n_{CNr}*r^2$ (gray dots) and $n_{CNr}*r$ (green dots) versus time. Note that the overlaid cometocentric distance r (blue line) is shown on a linear scale together with the sub-spacecraft latitude (red line), offset by -200 deg. 30 per cent total error bars on $n_{CNr}$ have been omitted for visual clarity.

Under the assumption of radial outflow at constant velocity and nuclear origin of the species, $n_{CNr}*r^2$, where r is the cometocentric distance of the spacecraft and hence of the observing instrument, should be proportional to the production rate. Fig. 4 (bottom) shows both the time-dependent variation of this value over the course of the Rosetta mission, as well as the corresponding cometocentric distance. It should be noted that a (usually) relatively small but variable portion of the CN signal may be due to outgassing and/or combustion from the spacecraft itself. While approaching the comet in early August 2014, the background density of CN was about 3000 cm$^{-3}$ and decreasing with time. The decreasing background leads to a decrease of $n_{CNr}*r^2$ in September 2014, which is not due to the cometary activity. As there is no reliable way to determine this background, it has not been subtracted from $n_{CNr}$ in Fig. 4 and thus introduces an additional source of error. As expected, the production rate proportional to $n_{CNr}*r^2$ peaks shortly after the comet's closest approach to the Sun in August 2015, when the outgassing from the cometary nucleus is generally maximal. While $\log(n_{CNr}*r^2)$ and r obviously correlate, $\log(n_{CNr}*r)$ as a function of time is comparably flat, suggesting that the outflow of $n_{CNr}$ might not decrease with $r^{-2}$. Variable observational parameters – the sub-spacecraft latitude is included in Fig. 4 (bottom) for comparison – and/or variable coma conditions may also affect the observations. A detailed investigation requires data wherein only the cometocentric distance varies while the rest of the observational parameters and coma conditions remain reasonably stable. A close flyby event on 14[th] February 2015 is therefore most suitable. On the outbound flyby trajectory, the sub-spacecraft latitude decreased continuously from 7.4° to -2.1°, while the phase angle increased continuously from 64.3° to 91.3°. Fig. 5 shows the counts measured for water ($c_{H2O}$), carbon dioxide ($c_{CO2}$), hydrogen cyanide ($c_{HCN}$) and the CN radical ($c_{CNr} = c_{CNt} - 0.168*c_{HCN}$). Deriving the dependence of the measured signal on the distance requires a coma model taking into account the varying latitude, longitude and phase angle and thus the inhomogeneous outgassing even on short time-scales. From a pure Haser model with homogeneous outgassing from a spherical nucleus, variation according to $r^{-2}$ is expected. A theoretical line representing $a*r^b+c$, where a is a scaling factor, b = -2 the decay factor expected from the Haser model and c the instrumental background, is included in Fig. 5. $c_{H2O}$ and $c_{CO2}$ show deviations from such an ideal case, most probably because of the nucleus' shape and the inhomogeneous outgassing. Modulations due to the nucleus rotations are clearly seen to differ for $c_{H2O}$ and $c_{CO2}$. The slope of $c_{HCN}$ is flatter, which may point to part of HCN also coming from an extended source. $c_{CNr}$ displays the flattest profile with modulations less pronounced than for $c_{H2O}$ or $c_{CO2}$. This would be expected if the CN radical is mainly released from a distributed source. The opposite latitudinal variation of $n_{CNr}*r^2$ and $n_{CNr}*r$ in Fig. 4 (bottom) e.g. around perihelion, as compared to the variations of the ratios in Fig. 2, 3 and 4 (top), seems to support a dust-related origin of CN: release of CN from dust grains would probably lead to a more isotropic distribution and hence to a less pronounced difference between the two hemispheres than for HCN.

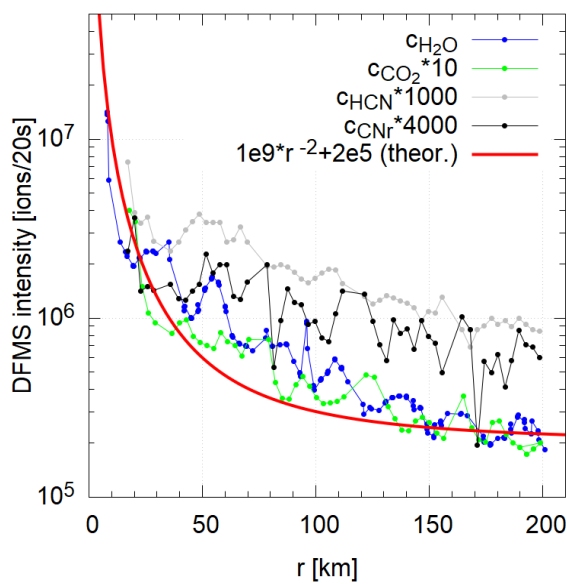

Figure 5. $c_{CNr}$ as compared to $c_{H2O}$, $c_{CO2}$ and $c_{HCN}$ observed by ROSINA/DFMS during a close flyby on 14th February 2015, as a function of the cometocentric distance r. The absolute values of $c_X$ have a statistical error of <10 per cent. However, they are also subject to a further 30 per cent systematic uncertainty. While this may shift the curves up or down, it should not change their slopes. Error bars have been omitted for visual clarity.

## 4. DISCUSSION

Data presented in subsections 3.1. and 3.2. show that ROSINA/DFMS observed a time-variable portion $c_{CNr}$ of the total CN signal $c_{CNt}$ that could not be explained by fragmentation of CN-bearing volatiles inside the instrument ($c_{CNf}$, mostly $c_{CNf-HCN}$). As shown in subsection 3.2., the corresponding local density $n_{CNr}$ is correlated with the distance of the orbiter from the cometary nucleus in a manner that suggests an at least partially distributed origin. This section is dedicated to the discussion of the data and their implications regarding the possible origin of the CN radical in particular.

### 4.1. Photodissociation of CN-bearing volatiles

In a cometary environment, the CN radical can form from the photodissociation of HCN (or other CN-bearing species), primarily under solar Ly-α irradiation. As stated in the introduction, prior to the reports of Bockelée-Morvan et al. (1984, 1985), this source was widely believed to completely account for the CN inventory of comets. Bodewits et al. (2016) observed the CN emission with the OSIRIS instrument's wide angle camera in the inner coma of 67P, finding both the brightness profile and the intensity to be inconsistent with CN originating solely from HCN photodissociation. However, the authors attributed the surplus to contamination by light emitted from $CO_2^+$ after ionizing electron impact excitation. The content of this subsection demonstrates for the case of the present study that the photodissociation of HCN can be safely neglected and that another point of origin must thus be sought.

For comet Hale-Bopp, Combi et al. (2000) estimated an optically thick coma with a radius of about 1000 km. 67P is much less active and its coma, even around perihelion, did not appear to be optically thick to Ly-α, cf. Shinnaka et al. (2017). HCN photodestruction rates have been estimated by Crovisier et al. (1994) to be 1.5e-5 s$^{-1}$ and by Huebner et al. (1992) to be 1.26e-5 s$^{-1}$ (quiet Sun conditions at 1 au). Using the CN photodestruction rate of 3.17e-6 s$^{-1}$ from Huebner et al. (1992) and neglecting HCN photoionization, the CN/HCN ratio due to photodissociation of HCN is estimated to be only 0.0013 at a cometocentric distance of 100 km during perihelion and even less earlier and later in the mission due to the shorter distances from the comet and larger heliocentric distances. The trends observed for the local CN densities relative to the local HCN densities in Fig. 4 (bottom) are not in agreement with trends expected from HCN photodissociation. This shows that despite the large quantum yield of 0.97 and under conditions similar to those encountered by 67P at its perihelion, HCN photodissociation plays a very minor role. CN radical production from the photodissociation of other CN-bearing volatiles is much lower than that from HCN for the simple reason of their significantly lower abundances, cf. Table 1. In a review on the topic, Fray et al. (2005) discussed potential CN parent molecules, based on the available photochemical data referenced therein. For acetonitrile ($CH_3CN$), a CN radical production about 100 times lower than that for HCN was estimated. For cyanoacetylene ($HC_3N$), the proposed quantum yield of 0.05 was very low. Both the photodissociation rate and the quantum yield of cyanogen ($C_2N_2$) is much disputed in the literature on the subject. The photodissociation rate of dicyanoacetylene ($C_4N_2$) is comparable to that of HCN, but no quantum yield is available. For hydrogen isocyanide (HNC), no photodissociation rate has been reported. The authors of the review concluded that only $HC_3N$ and $C_2N_2$ are viable candidates, together with HCN, for the reproduction of the observed CN profile. However, the remote observations that were reviewed pointed to a substantially lower abundance of $HC_3N$ than that which would be required, a result confirmed by the data in Table 1 and Rubin et al. (2019). $C_2N_2$ was not observed in comets until Altwegg et al. (2019) reported single detections in the coma of 67P, while $C_4N_2$ has not yet been identified. Consequently, neither of the observed CN-bearing species in 67P is abundant enough (apart from HCN) nor possesses the necessary photodissociation rate and quantum yield to contribute non-negligibly to the observed $n_{CNr}$.

### 4.2. Solar wind sputtering from CN-bearing refractories

CN radicals could also be formed by solar wind sputtering from cometary refractories. Wurz et al. (2015) stated that in the beginning of the Rosetta mission phase, the main solar wind surface access locations were on the southern hemisphere of the comet (local winter), because outgassing was dominated by the northern hemisphere (local summer) where solar wind was collisionally attenuated. The authors investigated the possibility of ion-induced sputtering (mostly by solar wind protons or alpha-particles) from the cometary surface. During the period under consideration (October 2015), the Rosetta spacecraft orbited the comet at a distance of about 10 km and the comet's activity was low, as it was still roughly 3.1 au from the Sun. The authors observed refractory elements (Na, K, Si and Ca) with ROSINA/DFMS, which were presumably sputtered from cometary refractories. Notably, no sputtered molecules had been observed. The sputtered atoms appeared to be anti-correlated with $H_2O$, a trend compatible with high $c_{CNt}/c_{HCN}$ over the southern hemisphere in Fig. 3 (top). If the observed $n_{CNr}$ is due to solar wind-induced sputtering from CN-bearing dust particles, its estimated abundance should increase as the comet approaches the Sun and decrease as soon as a magnetosheath begins to form (preventing solar wind access to the inner coma) and/or the inner coma becomes collisionally thick. The increased dust production around perihelion could counteract. To investigate the possibility of sputtering in detail, a complex analysis is needed, which exceeds the scope of this work. However, regarding the fact that ROSINA/DFMS could not detect any sputtered refractory elements anymore after October 2015 (Altwegg, priv. commun.), an efficient CN production due to sputtering around perihelion, as shown in Fig. 4 (bottom), seems unlikely.

### 4.3. Thermal degradation of CN-bearing refractories

As stated in the introduction, thermally degrading CN-bearing refractories have been widely discussed in the literature as additional (to HCN) and possibly distributed parents of the CN radical. Even though Woodney et al. (2002) reported a poor correlation of the observed CN with optical dust and a better correlation with HCN, many of these dust particles could actually be smaller than optical (micrometer-sized) dust. Very small CHON particles were first discovered by the mass spectrometers onboard the Giotto and Vega spacecrafts, as they flew past comet Halley in 1986. Their masses were reported by Kissel et al. (1989a,b) to be on the order of $10^{-16}$ g. However, their composition is not well understood. It is not possible to measure the composition of dust directly with ROSINA/DFMS, as the instrument was designed to measure volatiles. It is, however, still possible to detect the volatile products of degrading dust particles. Recently, Altwegg et al. (2017, 2020) published data collected from a dust grain, which, by chance, entered the ionization chamber of the DFMS during a period of high dust activity on 5th September 2016. The dust particle sublimated over a duration of approximately 2 hours, releasing many species that had previously been detected only in much lower relative abundances in the undisturbed coma and some which had never before been observed in situ at all. The authors reported, for instance, the detection of several products from the sublimation of ammonium salts, including those of ammonium cyanide ($NH_4CN$). Ammonium salts decompose mainly into $NH_3$ and the appropriate acid, e.g. HCN for $NH_4CN$. However, several other minor species were also detected in laboratory experiments by Hänni et al. (2019) and in space by Altwegg et al. (2020), reflecting the complexity of the sublimation process and raising the question of whether it may also be possible for ions and perhaps even radicals to form in relevant quantities under cometary conditions. Neither Hänni et al. (2019) nor Altwegg et al. (2020) provide a definitive answer and laboratory work on the topic continues.

The surface layer of 67P, which is reported to reach day-temperatures of e.g. about 230 K at 1.88 au from the Sun by Tosi et al. (2019), is thus probably depleted of most of its more volatile material (e.g. $NH_4CN$ sublimates at temperatures as low as 140 K according to Noble et al. (2013)). However, CN could be released from small dust particles ejected from deeper layers or from fresh surfaces. Small dust particles require less gas drag to be lifted up, such that even the gas drag present at the beginning of the mission, when 67P was still relatively far from the Sun, might already have sufficed. On its inbound trajectory, 67P was apparently covered with an accumulated dust mantle, cf. Schulz et al. (2015). Containing fallback material, as proposed e.g. by Keller et al. (2015), the dust layer in the North was generally thicker than the one in the South, but probably also more depleted in volatile and semi-volatile species. In the early mission phase, the less isolated southern hemisphere emitted much more $CO_2$ than the northern one, cf. Hässig et al. (2015). Sublimating $CO_2$, with a higher gas drag than water, could have lifted dust grains from the southern hemisphere. The thus released particles would then have heated up quickly to temperatures at which also less volatile material, as for instance the indicated ammonium salts, could have sublimated and/or thermally degraded. Such mechanism would be in agreement with both the overall trend of the $c_{CNt}/c_{HCN}$ ratios and the high values detected at the beginning of the mission especially over the southern hemisphere, see Fig. 2 and Fig. 3.

## 5. CONCLUSIONS

In this work, we presented the first in-situ data that has ever been available on cometary CN with unprecedented temporal and spatial resolution. We demonstrated that the total CN signal cannot be exhaustively explained by fragmentation of HCN inside ROSINA/DFMS, the high-resolution electron-impact mass spectrometer onboard the Rosetta orbiter. In a thorough analysis, we corrected for the instrument-inherent fragmentation effect by subtracting the portion of the CN signal produced from fragmenting CN-bearing volatile coma species (mainly HCN) and thus obtained the portion of the CN signal attributable to the CN radical. Data from a close flyby event in February 2015, when 67P was still about 2.3 au from the Sun, show that the cometocentric distance-dependence of CN does not agree with sublimation from the cometary nucleus alone. Based on our data, we discussed the following three scenarios for the origin of the CN observed in the inner coma of comet 67P: (1.) CN from photodissociation of CN-bearing molecules, especially HCN, which had the highest abundance. This does not produce sufficient CN to account for the amount observed, because even at cometocentric distances as small as about 100 km, less than one per cent of CN, relative to HCN, is estimated to have dissociated, even on 67P's closest approach to the Sun. Other CN-bearing candidate molecules, such as cyanogen ($C_2N_2$), cyanoacetylene ($HC_3N$) and acetonitrile ($CH_3CN$), are orders of magnitude less abundant than HCN and most of them also show considerably lower quantum yields. (2.) CN originating from solar wind sputtering from dust. This seems unlikely because the CN radical abundance does not obviously correlate with reported sputtering events. However, only a detailed model would be capable of suitably addressing the complexity of the matter. (3.) Thermal degradation of CN-bearing dust, leading to (partially) distributed CN production. A hypothetical scenario, wherein CN is produced from small lifted/ejected and subsequently heated cometary CN-bearing dust particles, is compatible with the observed latitudinal variation of the $c_{CN}/c_{HCN}$ ratio, as well as its temporal evolution over the two years of Rosetta's mission phase. $c_{CN}/c_{HCN}$ generally decreased on the inbound trajectory toward perihelion and slightly increased again on the outbound one, illustrating that the peak production rate of HCN is higher and its slope with heliocentric distance steeper than that of the observed additional source of CN. Such a dependence on the heliocentric distance might be observable remotely, based on data collected e.g. for comet Hale-Bopp between 4.6 au pre-perihelion and 12.8 au post-perihelion and for comet Hyakutake, cf. Rauer et al. (2003) and Fray et al. (2005) and literature referenced therein. However, both these comets belong to the group of four out of the eight comets featured in Fray et al. (2005) which exhibit similar production rates for HCN and the CN radical and which thus do not require an additional source to explain their CN abundances. For comets, however, in which the production rates of CN and HCN are known not to match, such observations have yet to be performed.

## Acknowledgements

ROSINA would not have produced such outstanding results without the work of the many engineers, technicians and scientists involved in the Rosetta mission and in the ROSINA instrument. Their contributions, as well as those of the entire ESA Rosetta team, are herewith gratefully acknowledged. Rosetta is an ESA mission with contributions from its member states and NASA. Work at the University of Bern was funded by the Canton of Bern and the Swiss National Science Foundation (200020 182418).
We would also like to express our thanks to Prof. P. Wurz and Dr. N. Fray for the stimulating discussions on the topics of sputtering and the cyano radical respectively.

## Data availability

The ROSINA/DFMS full mission data underlying this article are available in ESA's planetary science archive (PSA) at https://archives.esac.esa.int/psa.